\documentclass{article}

\usepackage{arxiv}

\usepackage[utf8]{inputenc} % allow utf-8 input
\usepackage[T1]{fontenc}    % use 8-bit T1 fonts
\usepackage{url}            % simple URL typesetting
\usepackage{amsfonts}       % blackboard math symbols
\usepackage{nicefrac}       % compact symbols for 1/2, etc.
\usepackage{microtype}      % microtypography

\usepackage{mathtools}
\DeclarePairedDelimiter{\ceil}{\lceil}{\rceil}
\usepackage{graphicx}

\usepackage{caption}

\usepackage{subcaption}

\usepackage[table,xcdraw]{xcolor}

\usepackage{amssymb}
\usepackage{textcomp}
\usepackage{xcolor}

\usepackage{algorithm} 
\usepackage{algpseudocode}

\usepackage{multirow}
\usepackage{cellspace}

\usepackage{hyperref}
\usepackage{doi}
\usepackage{eso-pic}
\newtheorem{prop}{Proposition}

\title{Impact of Differentials in SIMON$32$ Algorithm for Lightweight Security of Internet of Things}

\author{Jonathan Cook*,
        Sabih ur Rehman*
        and M. Arif Khan*\\ % <-this % stops a space\\
        {*School of Computing, Mathematics and Engineering, Charles Sturt University, Australia}}
\begin{document}
\maketitle

\begin{abstract}
	SIMON and SPECK were among the first efficient encryption algorithms introduced for resource-constrained applications. SIMON is suitable for Internet of Things (IoT) devices and has rapidly attracted the attention of the research community to understand its structure and analyse its security. To analyse the security of an encryption algorithm, researchers often employ cryptanalysis techniques. However, cryptanalysis is a resource and time-intensive task. To improve cryptanalysis efficiency, state-of-the-art research has proposed implementing heuristic search and sampling methods. Despite recent advances, the cryptanalysis of the SIMON cypher remains inefficient. Contributing factors are the large size of the difference distribution tables utilised in cryptanalysis and the scarcity of differentials with a high transition probability. To address these limitations, we introduce an analysis of differential properties of the SIMON$32$ cypher, revealing differential characteristics that pave the way for future efficiency enhancements. Our analysis has further increased the number of targeted rounds by identifying high probability differentials within a partial difference distribution table of the SIMON cypher, exceeding existing state-of-the-art benchmarks. The code designed for this work is available at https://github.com/johncook1979/simon32-analysis. 
\end{abstract}

% keywords can be removed

%%%%%%%%%%%%%%%%
%
% !!! - Generate bbl file from logs and outputs!!!
%
%%%%%%%%%%%%%%%%

\section{Introduction}  \label{Sec:Introduction}
Internet of Things (IoT) devices have traditionally exhibited limited processing capacity, hindering their ability to execute complex cyphers \cite{cook2023security}. Addressing the security gaps of unencrypted IoT data, in $2013$, the US National Security Agency (NSA) introduced SIMON, a lightweight cypher suitable for hardware-constrained applications, such as IoT \cite{beaulieu2015simon}. Since its introduction, the International Organisation for Standardisation (ISO) has provided support for block/key sizes greater than $64/96$ under ISO/IEC 29167-21:2018 \cite{iso2018simon}. The key to SIMON's strength lies in its series of transformation rounds that create confusion and diffusion to encrypt data \cite{singhal2021entropy}. However, while SIMON is a robust algorithm, it is crucial to analyse its security to ensure that the data encrypted by the algorithm remains secure. The imminent arrival of Sixth Generation (6G) networks necessitates an urgent need to ensure SIMON remains robust, as potential vulnerabilities can lead to compromised data and insecure devices. However, analysing and identifying vulnerabilities of SIMON$32$ requires using various cryptanalysis techniques without resorting to brute force methods \cite{de2006introduction, easttom2021cryptanalysis}. While Differential Cryptanalysis (DC) has demonstrated effectiveness against block algorithms, such as SIMON \cite{easttom2021cryptanalysis},  cryptanalysis itself remains inefficient. Many DC methods employ a Difference Distribution Table (DDT) \cite{biham1995matsui, beyne2022differential, heys2002tutorial}, which is a large matrix that allows researchers to analyse how differences to an input propagate to differences in the output. However, the size of DDTs can become vast \cite{biryukov2014automatic}. The DDT for a n-bit S-box, which normally operates on a 4-bit or 8-bit word, will comprise a matrix of $2^{n} \times 2^{n}$ \cite{biryukov2014automatic}. An S-box DDT for a small state size is relatively simple. However, addition, rotate XOR (ARX) cyphers such as SIMON, leverage modular addition to introduce non-linearity \cite{biryukov2014automatic}. Therefore, constructing a DDT for n-bit words is infeasible as it will require $2^{3n} \times 4$ bytes of memory for a standard $32$-bit word size \cite{biryukov2014automatic}. To address this limitation, the authors of \cite{biryukov2014automatic} proposed a partial DDT (pDDT) to reduce the search space. However, the distribution of significant differentials in a pDDT is sparse, which reduces the efficiency of cryptanalysis.

The three leading solutions to enhance efficiency are employing quantum computing \cite{anand2020evaluation}, which is not yet commercially available to consumers, applying heuristic search \cite{dwivedi2023security}, and reducing the search space through sampling \cite{cook2024lightweight}. While the heuristic may not produce optimal solutions to all problems, the answer will often be sufficient for the problem presented \cite{gigerenzer2011heuristic}. This ability to sufficiently solve complex problems establishes it as a viable solution for the cryptanalysis of lightweight encryption algorithms such as SIMON. However, as briefly mentioned above and explained in deeper detail in Section \ref{Subsec:pddt}, pDDTs contain numerous differentials with very few exhibiting a high probability of differential transition, highlighting a limitation of heuristics. For effective cryptanalysis, it is crucial to identify high-probability differential transitions. However, with a limited number of high-probability differential transitions in a pDDT, it is unlikely to discover them by chance. While the authors in \cite{cook2024lightweight} demonstrated that reducing the search space decreases the number of differentials and time required for cryptanalysis, heuristic methods remain largely ineffective. Heuristic techniques that rely on simple random selection may choose differentials that produce poor results, necessitating the selection of an alternative path, which increases time and computational resources. While existing work on SIMON$32$ presents substantial benchmarking through various techniques, many are inhibited by the size of the DDT and the method of optimal path self-discovery \cite{biryukov2014automatic, dwivedi2023security, cook2024lightweight}. This paper seeks to address these limitations by identifying optimal differential characteristics and patterns to facilitate efficient cryptanalysis of the SIMON cypher through an analytical investigation of differential characteristics.

Through a rigorous analysis of SIMON$32$ differentials, we discover and utilise distinct characteristics to enhance cryptanalysis efficiency. We expand on our contribution by generating the differential trails on the significant differentials, demonstrating an increase in the number of rounds targeted with improved probability over existing techniques. Although ISO standards pertain to block/key sizes of $64/96$ and greater, SIMON $16/32$ provides an experimental testbed to explore preliminary results which can be later applied to larger variants of SIMON. As stated by \cite{dwivedi2023security}, ideal cyphers follow the random distribution of cyphertext and plaintext pairs with a probability of $2^{2n}$, where $2n$ is the cyphers' size. If the differential path probability is greater than $2^{2n}$, the differential path can be used to distinguish the cypher from a random permutation. Previous studies reduced the number of rounds to achieve probabilities relative to the block size of the cypher. Increases in the number of rounds targeted without reducing the probability indicate improvement in the cryptanalysis technique. To understand SIMON$32$ differentials, this investigation takes a deep dive to identify differential characteristics to increase the number of targeted rounds efficiently. As this is an analytical study, our analysis is not negatively influenced by heuristic methods or 
seeks to be efficient.

\begin{figure}
    \centering
    \includegraphics[width=1\linewidth]{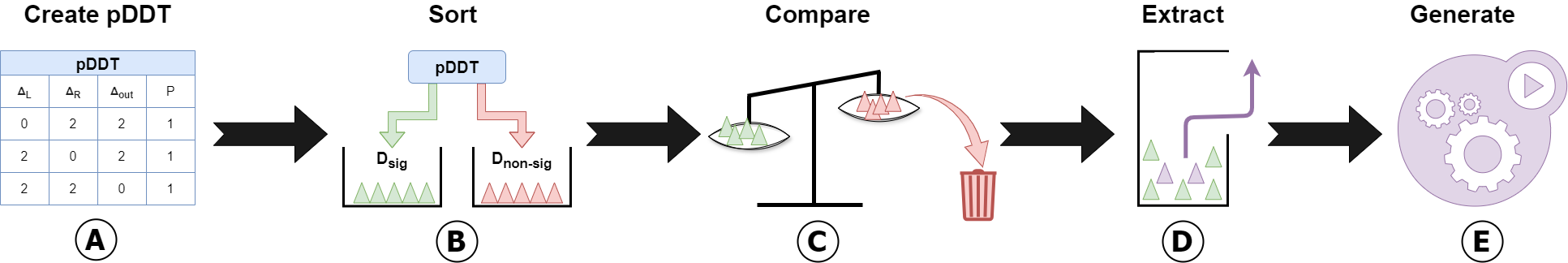}
    \caption{Our methodology }
    \label{fig:methodology}
\end{figure}

\section{Methodology \label{Sec:Methods}}

All experiments are performed on a standard retail Windows $10$ Enterprise edition $64$ bit laptop utilising Jupyter Notebook \cite{pimentel2021understanding}. The methodology is summarised in Figure \ref{fig:methodology}, which describes the processes employed in our investigation. We begin at step \textit{A} by creating a pDDT. Differentials are sorted at step \textit{B} based on their calculated probability. At \textit{C}, we conduct preliminary experiments and compare Hamming Weight (HW) distributions. At step \textit{D}, we extract the promising differentials from the preliminary experiments for further analysis and evaluate their differential trails and cumulative probability over $20$ rounds (Step \textit{E}).

\subsection{Create pDDT}\label{Subsec:pddt}

A DDT for the SIMON$32$ cypher is infeasibly large and time-consuming to produce. To address the large size of DDTs, \cite{biryukov2014automatic} proposed the creation of a partial DDT (pDDT) where only differences above a Differential Probability (DP) threshold ($P_{thr}$) are recorded. The pDDT can be expressed as:

\begin{equation} 
    (a,b,c) \in pDDT \Leftrightarrow DP(a, b \rightarrow c) \geq P_{thr}.
\end{equation}

The efficient computation of the pDDT can be executed with the following proposition: 

\vspace{2mm}
\begin{prop}
The DP of the XOR operation combined with addition modulo $2n$ decreases as the word size of the differences $a$, $b$, $c$ increases, such that:

\begin{equation} 
    p_{n} \leq \dots \leq p_{k} \leq p_{k-1} \leq \dots \leq p_{1} \leq p_{0},
\end{equation}
\end{prop}

where $p_k$ denotes the DP from $a_k$, $b_k$ progressing to $c_k$, where $n \geq k \geq 1,p_0=1$ and $x_k$ denotes the $k$ Least-Significant Bit (LSB) difference $x$, such that $x_k=x[k-1:0]$. The algorithm produces the pDDT commencing at the LSB position $k=0$. It then allocates the values to $a_k$, $b_k$ and $c_k$ recursively \cite{biryukov2014automatic}. At each bit position $k$, for $n > k \geq 0$, the likelihood of a $(k + 1)^{th}$ bit differential with a partial structure is evaluated to establish whether it exceeds the specified threshold. If true, the algorithm proceeds to the next bit. If not, it returns and assigns different values to $a_k$, $b_k$ and $c_k$. $k$ is initialised with a value of $0$, such that $a_0 = b_0 = c_0 = \Phi$. The algorithm is repeated until $k=n$, at which point it is terminated and the XOR differential $(a_k,~b_k~ \rightarrow ~ c_k)$ is added to the pDDT \cite{biryukov2014automatic}. We define our threshold as $0.1$ in Algorithm \ref{Alg:compute_pddd}, which produces $3951388$ pDDT elements. Reducing the threshold increases the number of pDDT elements and generation time.

\begin{algorithm}
    \small
	\caption{: Compute pDDT}
    \label{Alg:compute_pddd}
    \hspace*{\algorithmicindent} \textbf{Input}: \texttt{$n,~p_{thr},~k,~p_k,~a_k,~b_k,~c_k$} \\
    \hspace*{\algorithmicindent} \textbf{Output}: \texttt{ pDDT D:(a, b, c) $\in$ D:} \\ 
    \hspace*{\algorithmicindent}\hspace{\algorithmicindent}\hspace{\algorithmicindent}\hspace{\algorithmicindent} $DP(a,b \rightarrow c) \geq p_{thr}$  
	\begin{algorithmic}[1]
    \Function{calcpddt}{$n,~p_{thr},~k,~p_k,~a_k,~b_k,~c_k$ }
        \If{\texttt{n} == \texttt{k}}
            \State Add ($a,~b,~c$)$ \leftarrow $($a_k,~b_k,~c_k$) \texttt{to D} 
        \EndIf \\
        \hspace{\algorithmicindent} \Return 
        \For{x,y,z $\in 0,1$}
            \State $a_k+1 \leftarrow x \mid a_k,b_k+1 \leftarrow y \mid b_k,c_k+1 \leftarrow z \mid c_k$
            \State $p_k + 1 = DP(a_k + 1, b_k + 1 \rightarrow c_k + 1)$
            \If{$p_k + 1 \geq p_{thr}$}
                \State \texttt{CALCPDDT}($n,~p_{thr},~k+1,~p_k+1,~a_k+1,~ b_k+1,~c_k+1$)
            \EndIf
        \EndFor \\
        \hspace{\algorithmicindent} \Return
    \EndFunction
 
	\end{algorithmic} 
\end{algorithm}

\subsection{Sorting differentials}
To determine if the differentials are statistically significant, they are sorted into two sets of data.
The sorting of differentials $D$, representing the set of differentials, is presented in Algorithm \ref{Alg:filterDiffs} and is expressed as:
\begin{equation}
    D = \{d_1, d_2, \dots d_n\},
\end{equation}
\indent where $d$ represents each differential up to $n$ differentials. Each $d_n$ has an associated probability used to sort significant differentials ($D_{sig}$), where the DP is equal to or greater than $0.5$, and non-significant differentials $D_{non-sig}$ as: 

\begin{equation}
    D_{sig} = d_n \in D \mid P_{d_n} \geq P_{thr},
\end{equation}
\begin{equation}
    D_{non-sig} = d_n \in D \mid P_{d_n} < P_{thr},
\end{equation}

\indent where $P(d_n)$ is the differential probability and $P_{thr}$ is the predefined probability threshold for sorting. While sorting creates two subsets of data, the number of insignificant differentials is quantifiably higher. Therefore, it is infeasible to test each non-significant differential as it will consume significant resources. Therefore, we take a quota-based sample of $10$\% of the total insignificant output differentials for analysis \cite{cook2024lightweight}. The number of samples to take from the population of $n$ is given as:

\begin{equation} 
    n = \text{max} \left( 1, \ceil*{\left(\frac{x}{100} \times N_h\right)} \right),
    \label{eq:nonSigSam}
\end{equation}
\indent where $x$ is the sample size of the stratum, and $N_h$ is the total number of observations of the $h^{th}$ type in the population, with at least one of each $h^{th}$ discrete value, and $\ceil*{ . }$ represents the ceiling function.

% Filter differentials
\begin{algorithm}[t]
\small
	\caption{: Sort differentials based on probability}
    \label{Alg:filterDiffs}
    \hspace*{\algorithmicindent} \textbf{Input}: \texttt{pDDT} \\
    \hspace*{\algorithmicindent} \textbf{Output}: Significant differentials list as \texttt{sigDif}\\
    \hspace*{\algorithmicindent} \textbf{Output}: Non-significant differentials list as \texttt{nonSigDif}
	\begin{algorithmic}[1]
    \State Initialise array of significant differentials as \texttt{sigDif}
    \State Initialise array of non-significant differentials as \texttt{nonSigDif}
    \For{each \texttt{differential} in \texttt{PDDT}}
        \If{\texttt{differential} $\geq$ $0.5$}
            \State Append to \texttt{sigDif}
        \Else 
            \State Append to \texttt{nonSigDif}
        \EndIf
    \EndFor 
	\end{algorithmic} 
\end{algorithm}

\subsection{Compare hamming weight distribution}\label{Subsec:CompHW}
Following sorting, we perform a preliminary analysis of HW distribution.
We take a sample of non-significant differentials, extracted using Equation \ref{eq:nonSigSam}, and significant differentials for analysis. The HW of each set of differentials is calculated employing Algorithm \ref{Alg:ExecuteExperiments}. The distributions of HWs are analysed and compared. Differentials that produce a zero HW indicates that the input pair did not result in any observable output difference. Algorithm \ref{Alg:ExecuteExperiments} can be called any number of times to calculate differential HWs.

% Main experiment function. Can be called when needed to execute experiments
\begin{algorithm}[t]
\small
	\caption{: Execute experiments}
    \label{Alg:ExecuteExperiments}
    \hspace*{\algorithmicindent} \textbf{Input}: \texttt{differential}, \texttt{num\_trials} \\
    \hspace*{\algorithmicindent} \textbf{Output}: Array of hamming weights as \texttt{results}
	\begin{algorithmic}[1]
    \Function{ExecuteExperiment}{}
        \State Initialise \texttt{mask} = $2^{16} - 1$
        \State Initialise \texttt{alpha}, \texttt{beta}, \texttt{gamma} = 1, 8, 2
        \State Initialise \texttt{SIMON\_ROUNDS} = $10$
        \State Initialise empty array \texttt{results} = []
        \For{each trial from 1 to \texttt{num\_trials}}
            \State Calculate differential hamming weight
            % \State Set \texttt{st$1$} to $\triangle$X
            % \State Set \texttt{st$2$} to $\triangle$Y
            % \State Weight = \Call{FindPath}{st$1$, st$0$, \texttt{SIMON\_ROUNDS}, \texttt{mask}, \texttt{alpha}, \texttt{beta}, \texttt{gamma}}
            \State Append \texttt{Weight} to \texttt{results}
        \EndFor \\
        \hspace{\algorithmicindent}\Return \texttt{results}
    \EndFunction
	\end{algorithmic} 
\end{algorithm}

\subsection{Extract promising differentials}
Significant differentials that produce a near zero HW are extracted for further analysis. This step involves extracting the most promising differentials from the set of significant differentials. The extraction of promising differentials $D_{prom}$ is defined as:

\begin{equation}
    D_{prom} = d_n \in D_{sig} \mid HW_{d_n} = 0,
\end{equation}
\indent where $HW_{d_n}$ is the differential HW.

\subsection{Generate differential trails}
Once the most promising differentials have been identified, the next stage is to test and generate differential trails. This will yield the HW at each round and the DC probability. The round simulation is presented in Algorithm \ref{Alg:ID0Diffs} and is called for each of the promising differentials. Algorithm \ref{Alg:ID0Diffs} calls the simulate rounds function presented in Algorithm \ref{Alg:SimRounds}. The number of SIMON$32$ rounds is passed as a function attribute. Algorithm \ref{Alg:SimRounds} returns each round $\triangle_L$, $\triangle_R$ and logarithmic probability as $log_2P$. Algorithm \ref{Alg:ID0Diffs} returns the results with the sum of the round $r$ probability for the differential as $\sum_rP_r$.

% Significant differential trails
% \enlargethispage{0.2in}
\begin{algorithm}
\small
	\caption{: Generate differential trails of promising differentials}
    \label{Alg:ID0Diffs}
    \hspace*{\algorithmicindent} \textbf{Input}: Significant Differentials as \texttt{sigDif} \\
    \hspace*{\algorithmicindent} \textbf{Output}: Differential trails
	\begin{algorithmic}[1]
    \State SIMON\_ROUNDS = 18
    \For{\texttt{diff} in \texttt{sigDif}}
        \State differentials, probabilities = \Call{SimRounds}{\texttt{diff, SIMON\_ROUNDS}}
        \For{Index, differential, probability in (differentials, probabilities)}
            \State $\triangle_L$, $\triangle_R$, $\triangle_{out}$ = differential \\
           \hspace{\algorithmicindent}\hspace{\algorithmicindent}\Return $\triangle_L$, $\triangle_R$, \texttt{probability}
        \EndFor
    \EndFor
	\end{algorithmic} 
\end{algorithm}

% Simulate rounds
\begin{algorithm}
\small
	\caption{: Simulate rounds}
    \label{Alg:SimRounds}
    \hspace*{\algorithmicindent} \textbf{Input}: \texttt{differential, SIMON\_ROUNDS}\\
    \hspace*{\algorithmicindent} \textbf{Output}: Array of differentials as \texttt{differentials} \\
    \hspace*{\algorithmicindent} \textbf{Output}: Array of probabilities as \texttt{probabilities}
	\begin{algorithmic}[1]
    \Function{SimRounds}{}
        %\State Initialise \texttt{SIMON\_ROUNDS} = $10$
        \State Initialise empty array \texttt{differentials} = []
        \State Initialise empty array \texttt{probabilities} = []
        \For{each round in \texttt{SIMON\_ROUNDS}}
            \State $\triangle_L$, $\triangle_R$, $\triangle_{out}$ = Calculated SIMON round
            \State \texttt{HW} = Calculate HW
            \State \texttt{probability} = \texttt{2**(-HW)} 
            \State Append \texttt{probability} to \texttt{probabilities}
            \State Append \texttt{$\triangle_L$, $\triangle_R$, $\triangle_{out}$} to \texttt{differentials}
        \EndFor \\
        \hspace{\algorithmicindent}\Return \texttt{differentials}\\
        \hspace{\algorithmicindent}\Return \texttt{probabilities}
        
    \EndFunction
	\end{algorithmic} 
\end{algorithm}

\section{Results and Analysis \label{Sec:Results}}

This section presents the results of our experiments to identify significant differentials for the cryptanalysis of the SIMON$32$ cypher. 
We begin by analysing the distribution of differentials between two data subsets to identify potential high-probability differentials. We then extract these differentials and examine their trails and associated probabilities. Finally, the results are sorted to highlight the differentials with the most promising probability transitions.

\begin{figure}
    \centering
    \includegraphics[width=0.9\linewidth]{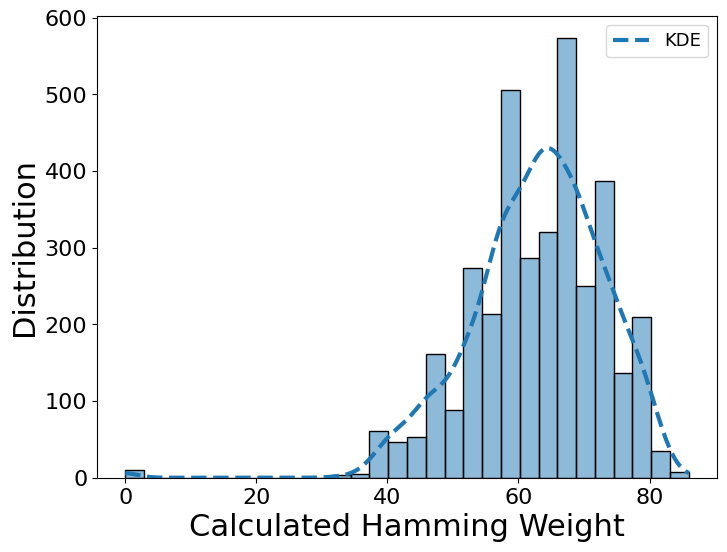}
    \caption{Frequency distribution of significant differentials}
    \label{fig:significant_hw_dist}
\end{figure}

\begin{figure}
    \centering
    \includegraphics[width=0.9\linewidth]{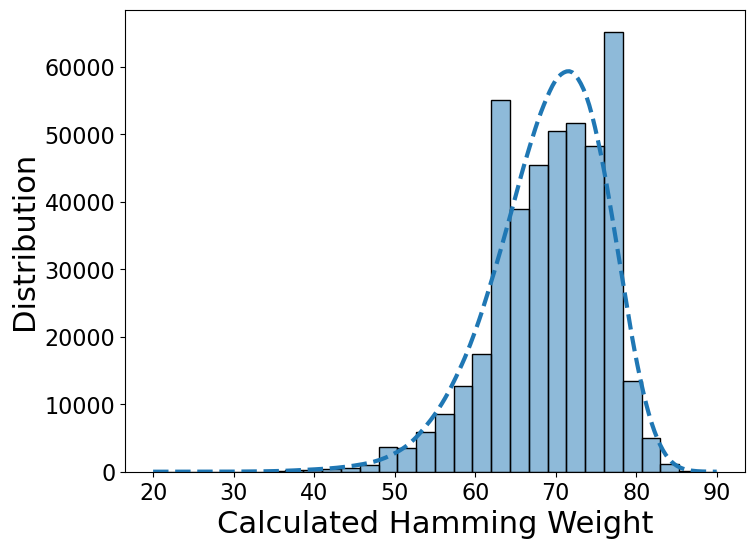}
    \caption{Frequency distribution of non-significant differentials}
    \label{fig:non_significant_hw_dist}
\end{figure}

\begin{figure}
    \centering
    \includegraphics[width=0.9\columnwidth]{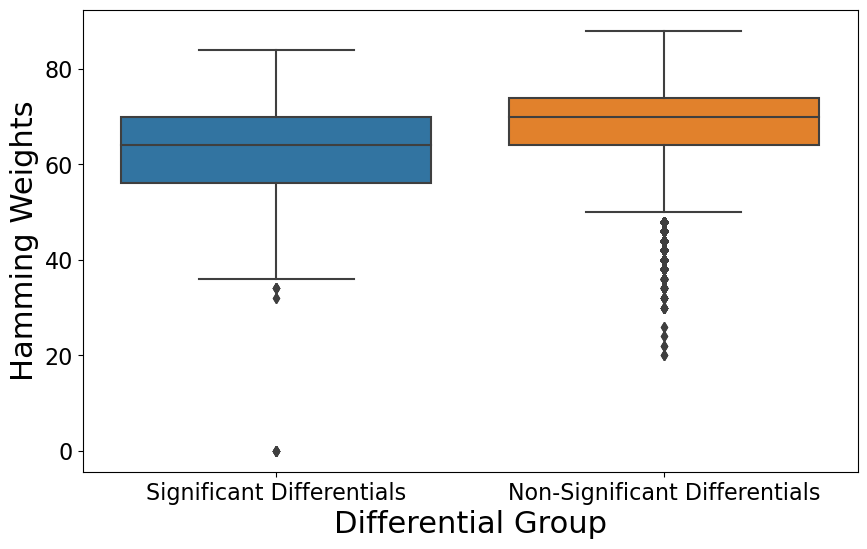}
    \caption{Boxplot of differential Hamming Weights}
    \label{fig:differential_hw_boxplot}
\end{figure}

\begin{figure*}
    \centering
    \includegraphics[width=0.8\linewidth]{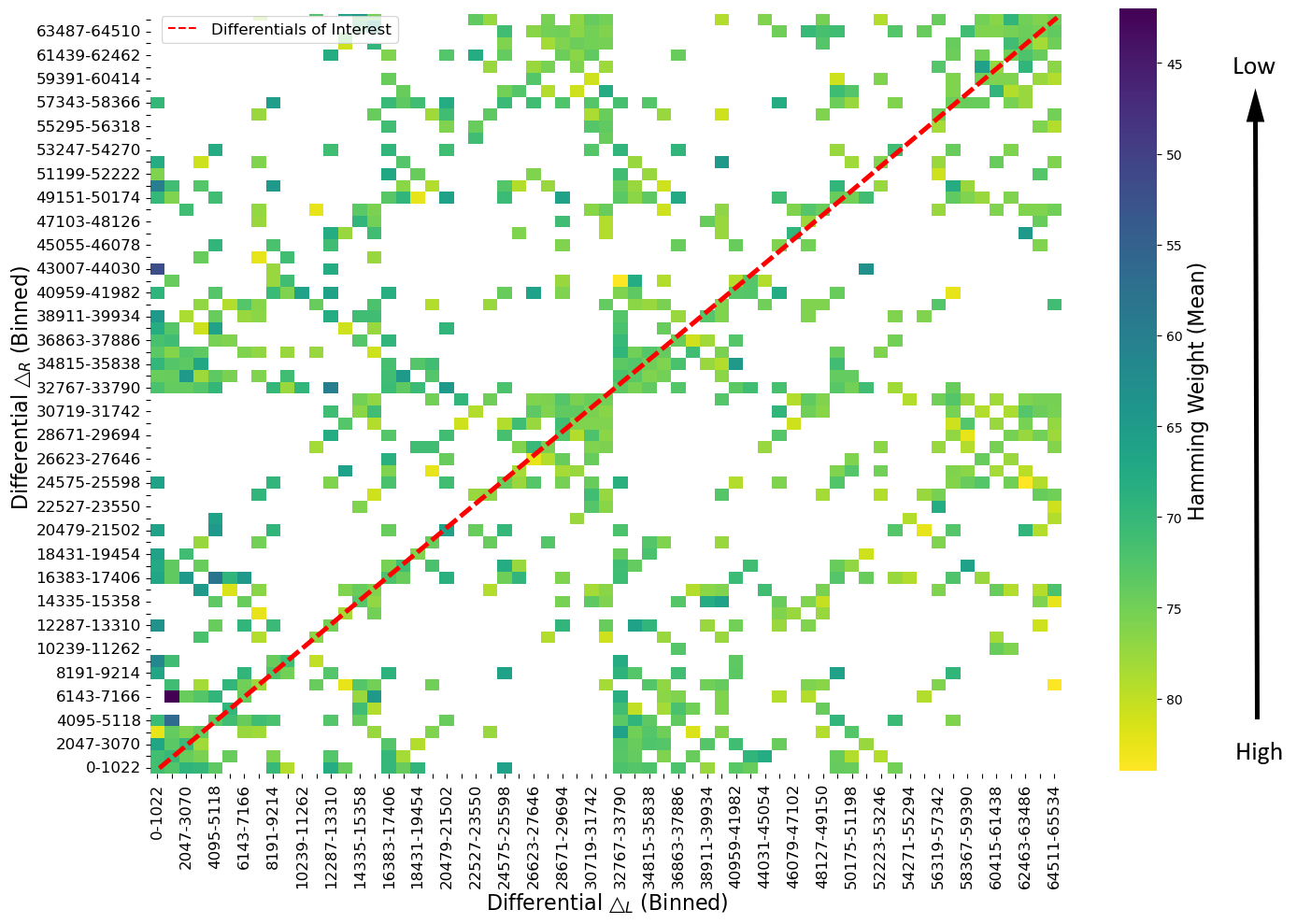}
    \caption{Heat map of mean best hamming weight grouped by input differentials}
    \label{fig:heatmap}
\end{figure*}

As illustrated in Figures \ref{fig:significant_hw_dist} and \ref{fig:non_significant_hw_dist}, the significant and non-significant differentials produce a similar distribution of HWs. However, as evidenced in Figure \ref{fig:significant_hw_dist}, some significant differentials produced a HW near $0$ while non-significant differentials did not. Additionally, an observation of the distribution of both plots indicates a binomial distribution of HWs \cite{wehrspohn2022take}. However, both figures are slightly left-skewed, indicating a median higher HW. Further investigations on finding the exact distribution of the data is the future work of this paper. A HW of $0$ indicates that there are transitions within some differentials where there is no difference to the output. As shown in Figure \ref{fig:differential_hw_boxplot}, significant differentials indicate the presence of differentials that produce a HW near $0$, while the non-significant differentials do not. Although outliers in the non-significant differentials are present, they're not near $0$.

We can examine the HW distribution relative to input differentials in greater detail using a heatmap, as presented in Figure \ref{fig:heatmap}. We grouped the differentials into smaller bins by dividing the $16$ bit space into $64$ equal parts. Each cell represents the mean HW over $10$ simulated SIMON$32$ rounds of all differentials within the range. An interesting observation is the linear convergence of differentials along the $45$ degree gradient highlighted by the red dashed line. As shown, powers of two differentials reside within the bins along the same gradient. Although these bins do not produce the lowest mean HW, influences of other differentials within the bin alter the mean HW.

As highlighted by the colour intensity, most differentials produce a HW between $65$ and $75$, with very few producing low HWs. Indeed, from the $4096$ cells, only $2$ produce a significantly low mean HW. Interestingly, cells that align with power of two differentials do not produce the deepest colour. This suggests the mean HW of differential transitions within the bin is skewed. We can scrutinise these results by extracting differentials within one of the aligned bins and evaluating the sum of their logarithmic probability over $10$ rounds. We take the convergence of the cells within the centre of the heatmap, aligning with the power of two differential of $32768$. While the mean HW suggests a value of approximately $70$, calling Algorithm \ref{Alg:SimRounds} on the differentials $32768$ within the bin produces a summed logarithmic probability of $2^{-17}$, far below the mean results within the bin. This indicates that the mean value is skewed by non-significant differentials. Furthermore, although several cells produce a light colour, the bins are largely devoid of power of two differentials, with fewer non-significant differentials to skew the mean result. Finally, as observed by the colourbar legend, the lowest mean HW is $30$, higher than the powers of two differentials tested above. This indicates an abundance of non-significant differentials influencing results.

Next, we apply statistical testing to validate our observations in Figures \ref{fig:significant_hw_dist}, \ref{fig:non_significant_hw_dist} and \ref{fig:differential_hw_boxplot}. We employ the p-value which is a statistical method used in hypothesis testing to determine the significance of the result \cite{jawlik2016statistics}. In addition to the p-value, we employ the t-statistic commonly used in t-tests comparing the means of two groups. We first analyse the significant and non-significant differentials to evaluate the p-value. For our analysis, the p-value returned $0.0$, indicating a very low probability of the observed differences occurring by chance. To validate the results, the t-statistic returned a value of $-65.09$, suggesting that the mean between the two is large. This validates the p-value, indicating a high degree of statistically significant differences between the two subsets of differentials.

Next, we extract $31$ promising differentials from the full set of significant differentials for further investigation. An observation of the promising differentials reveals that all output differentials are a value of $0$, and most input differentials are a power of two. As presented in Table \ref{tab:diff_trail}, the best differential trail identified from the promising differentials has a probability of $2^{-32}$ from a total of $20$ simulated SIMON$32$ rounds, exceeding previous best results as presented in Table \ref{tab:comparison}. While SIMON is predominantly a bitwise encryption algorithm represented in binary form, analysis of differential propagation is commonly undertaken using hexadecimal format. As highlighted in Section \ref{Sec:Introduction}, attaining a differential path probability equal to or greater than the size of the cypher distinguishes the path from random permutation. Previous investigations achieved similar results over fewer rounds. Indeed, both \cite{cook2024lightweight} and \cite{dwivedi2023security} acknowledged that increasing the number of rounds significantly hampered their heuristic efficiency.

\begin{table}[ht]
\centering
\tiny
\caption{Best significant differential trail}
\label{tab:diff_trail}
\begin{tabular}{|l|l|l|r|}
\hline
\textbf{Round} & \textbf{~$\Delta _L$} & \textbf{$\Delta _R$} & \multicolumn{1}{l|}{\textbf{$log_2P$}} \\ \hline
$0$            & ~$0xa000$             & $0x8000$             & $-1$                                   \\ \hline
$1$            & ~$0xa800$             & $0xa000$             & $-1$                                   \\ \hline
$2$            & ~$0x8a00$             & $0xa800$             & $-2$                                   \\ \hline
$3$            & ~$0x8a80$             & $0x8a00$             & $-1$                                   \\ \hline
$4$            & ~$0xa8a0$             & $0x8a80$             & $-3$                                   \\ \hline
$5$            & ~$0xa0a8$             & $0xa8a0$             & $-2$                                   \\ \hline
$6$            & ~$0x808a$             & $0xa0a8$             & $-3$                                   \\ \hline
$7$            & ~$0x8a$               & $0x808a$             & $-1$                                   \\ \hline
$8$            & ~$0xa8$               & $0x8a$               & $-2$                                   \\ \hline
$9$            & ~$0xa0$               & $0xa8$               & $-1$                                   \\ \hline
$10$           & ~$0x80$               & $0xa0$               & $-1$                                   \\ \hline
$11$           & ~$0x80$               & $0x80$               & $0$                                    \\ \hline
$12$           & ~$0xa0$               & $0x80$               & $-1$                                   \\ \hline
$13$           & ~$0xa8$               & $0xa0$               & $-1$                                   \\ \hline
$14$           & ~$0x8a$               & $0xa8$               & $-2$                                   \\ \hline
$15$           & ~$0x808a$             & $0x8a$               & $-1$                                   \\ \hline
$16$           & ~$0xa0a8$             & $0x808a$             & $-3$                                   \\ \hline
$17$           & ~$0xa8a0$             & $0xa0a8$             & $-2$                                   \\ \hline
$18$           & ~$0x8a80$             & $0xa8a0$             & $-3$                                   \\ \hline
$19$           & ~$0x8a00$             & $0x8a80$             & $-1$                                   \\ \hline
$\sum_rP_r$    & ~~                    & ~                    & $-32$                                  \\ \hline
\end{tabular}
\end{table}

% \usepackage{multirow}+
% \setlength{\cellspacetoplimit}{2pt}
% \setlength{\cellspacebottomlimit}{2pt}
\begin{table}[ht]
\centering
\caption{Summary of differential paths for SIMON.}
\label{tab:comparison}
% \resizebox{6.5cm}{!}{%
\tiny
\begin{tabular}{|Sl|Sc|Sc|Sc|Sc|}
\hline
\multicolumn{1}{|c|}{\textbf{Cypher}} & \begin{tabular}[c]{@{}l@{}}Distingisher\\ rounds\end{tabular} & \textbf{Probability} & \textbf{Year} & \textbf{Reference}           \\ \hline
% \multicolumn{1}{|c|}{\textbf{Cypher}} & \textbf{Distinguisher rounds} & \textbf{Probability} & \textbf{Year} & \textbf{Reference}           \\ \hline
\multirow{7}{*}{SIMON$32$}                & $13$                          & $2^{-29.69}$         & $2009$        & \cite{cazenave2009nested}    \\ \cline{2-5} 
                                      & $13$                          & $2^{-30.2}$          & $2015$        & \cite{abed2015differential}  \\ \cline{2-5} 
                                      & $14$                          & $2^{-30.81}$         & $2015$        & \cite{kolbl2015observations} \\ \cline{2-5} 
                                      & $11$                          & $2^{-30.0}$          & $2017$        & \cite{liu2017optimal}        \\ \cline{2-5} 
                                      & $15$                          & $2^{-32.0}$          & $2023$        & \cite{dwivedi2023security}   \\ \cline{2-5} 
                                      & $9$                          & $2^{-30.82}$             & $2024$        & \cite{qiao2024bit}   \\
                                      \cline{2-5}
                                      & $17$                          & $2^{-68.83}$             & $2024$        & \cite{qiao2024bit}   \\
                                      \cline{2-5}
                                      & $15$                          & $2^{-32.0}$          & $2024$        & \cite{cook2024lightweight}   \\ \cline{2-5} 
                                      
                                      & \cellcolor[HTML]{dcd5d5}$20$                          & \cellcolor[HTML]{dcd5d5}$2^{-32.0}$          & \cellcolor[HTML]{dcd5d5} $2025$        & \cellcolor[HTML]{dcd5d5}Our work                   \\ \hline 
\end{tabular}
% }
\end{table}

Examining the results of this analysis highlights several interesting cypher characteristics. The prevalence of power of two differentials and the limited bitwise movements and permutations of the cypher rounds highlight potential future avenues of investigation. 
The features identified can be utilised to enhance differential cryptanalysis efficiency.
Although the results presented in this work open unexplored avenues for the effective cryptanalysis of the SIMON$32$ cypher, it is not an exhaustive investigation. 
Nevertheless, the investigation identifies differentials that exceed the number of rounds while maintaining a low probability.

\section{Conclusion \label{Sec:Conclusion}}
Current differential cryptanalysis techniques targeting the SIMON$32$ cypher, such as heuristics and sampling, are inefficient large search space and a high distribution of low-probability differentials. To address existing limitations, this paper presented a technique based on an analytical investigation of differentials to identify optimal paths from a partial difference distribution table. From these optimal differentials, we have identified specific characteristics, such as power of two differentials, which can be exploited to narrow the search space and improve differential cryptanalysis efficiency. While not all identified differentials produce desirable optimal results, the best optimal results in this paper exceed the best results in previous works. Although this investigation is focused on the SIMON$32$ variant as a proof of concept, the analytical investigation presented in this paper can be scaled to other variants of the SIMON family and SIMON-like cyphers. 

Furthermore, future studies can leverage contemporary disciplines such as machine learning and knowledge graphs to exploit differential characteristics, improving cryptanalysis efficiency and increasing the number of targeted rounds. Finally, investigating power of two properties in the cryptanalysis of the SIMON cypher can yield further efficiency enhancements.

 \nocite{*} 
\bibliographystyle{unsrt}
\bibliography{references}  %%% Uncomment this line and comment out the ``thebibliography'' section below to use the external .bib file (using bibtex) .

@inproceedings{beaulieu2015simon,
  title={{The SIMON and SPECK lightweight block ciphers}},
  author={Beaulieu, Ray and Shors, Douglas and Smith, Jason and Treatman-Clark, Stefan and Weeks, Bryan and Wingers, Louis},
  booktitle={{Proceedings of the 52nd Annual Design Automation Conference}},
  pages={1--6},
  year={2015}
}

@article{de2006introduction,
  title={{An introduction to Block Cipher Cryptanalysis}},
  author={De Canniere, Christophe and Biryukov, Alex and Preneel, Bart},
  journal={Proceedings of the IEEE},
  volume={94},
  number={2},
  pages={346--356},
  year={2006},
  publisher={IEEE}
}

@article{easttom2021cryptanalysis,
  title={Cryptanalysis},
  author={Easttom, William and Easttom, William},
  journal={Modern Cryptography: Applied Mathematics for Encryption and Information Security},
  pages={357--372},
  year={2021},
  publisher={Springer}
}

@article{dwivedi2023security,
  title={{Security analysis of lightweight IoT encryption algorithms: SIMON and SIMECK}},
  author={Dwivedi, Ashutosh Dhar and Srivastava, Gautam},
  journal={Internet of Things},
  pages={100677},
  year={2023},
  publisher={Elsevier}
}

@inproceedings{biryukov2014automatic,
  title={{Automatic Search for Differential Trails in ARX Ciphers}},
  author={Biryukov, Alex and Velichkov, Vesselin},
  booktitle={{Proceedings of The Cryptographer’s Track at the RSA Conference: Topics in Cryptology--CT-RSA }},
  pages={227--250},
  year={2014},
  organization={Springer}
}

@inproceedings{abed2015differential,
  title={{Differential Cryptanalysis of Round-Reduced Simon and Speck}},
  author={Abed, Farzaneh and List, Eik and Lucks, Stefan and Wenzel, Jakob},
  booktitle={Proceedings of 21st International Workshop Fast Software Encryption (FSE)},
  pages={525--545},
  year={2015},
  organization={Springer}
}

@inproceedings{cazenave2009nested,
  title={{Nested Monte-Carlo Search}},
  author={Cazenave, Tristan},
  booktitle={{Proceedinsg of 21st International Joint Conference on Artificial Intelligence}},
  volume={9},
  pages={456--461},
  year={2009}
}

@inproceedings{kolbl2015observations,
  title={{Observations on the SIMON Block Cipher Family}},
  author={K{\"o}lbl, Stefan and Leander, Gregor and Tiessen, Tyge},
  booktitle={{Proccedings of the 35th Annual Cryptology Conference: Advances in Cryptology--CRYPTO 2015}},
  pages={161--185},
  year={2015},
  organization={Springer}
}

@article{cook2024lightweight,
  title={{Lightweight Cryptanalysis of IoT Encryption Algorithms: Is Quota Sampling the Answer?}},
  author={Cook, Jonathan and Ur Rehman, Sabih and Khan, M Arif},
  journal={IEEE Access},
  year={2024},
  publisher={IEEE}
}

@article{gigerenzer2011heuristic,
  title={{Heuristic Decision Making}},
  author={Gigerenzer, Gerd and Gaissmaier, Wolfgang},
  journal={Annual Review of Psychology},
  volume={62},
  pages={451--482},
  year={2011},
  publisher={Annual Reviews}
}

@inproceedings{beyne2022differential,
  title={{Differential Cryptanalysis in the Fixed-Key Model}},
  author={Beyne, Tim and Rijmen, Vincent},
  booktitle={{Proceedings of Advances in Cryptology -- CRYPTO 2022}},
  pages={687--716},
  year={2022},
  organization={Springer}
}

@article{heys2002tutorial,
  title={{A Tutorial on Linear and Differential Cryptanalysis}},
  author={Heys, Howard M},
  journal={Cryptologia},
  volume={26},
  number={3},
  pages={189--221},
  year={2002},
  publisher={Taylor \& Francis}
}

@article{singhal2021entropy,
  title={{Entropy Reduction Model for Pinpointing Differential Fault Analysis on SIMON and SIMECK Ciphers}},
  author={Singhal, Naman and Joshi, Priyanka and Mazumdar, Bodhisatwa},
  journal={IEEE Transactions on Computer-Aided Design of Integrated Circuits and Systems},
  volume={40},
  number={6},
  pages={1090--1101},
  year={2021},
  publisher={IEEE}
}

@article{liu2017optimal,
  title={{Optimal Differential Trails in SIMON-like Ciphers}},
  author={Liu, Zhengbin and Li, Yongqiang and Wang, Mingsheng},
  journal={Cryptology ePrint Archive},
  year={2017}
}

@book{jawlik2016statistics,
  title={{Statistics from A to Z: Confusing Concepts Clarified}},
  author={Jawlik, Andrew A},
  year={2016},
  publisher={John Wiley \& Sons}
}

@inproceedings{anand2020evaluation,
  title={{Evaluation of Quantum Cryptanalysis on SPECK}},
  author={Anand, Ravi and Maitra, Arpita and Mukhopadhyay, Sourav},
  booktitle={{Proceedings of 21st International Conference on Cryptology in India, Progress in Cryptology--INDOCRYPT 2020}},
  pages={395--413},
  year={2020},
  organization={Springer}
}

@article{cook2023security,
  title={{Security and Privacy for Low Power IoT Devices on 5G and Beyond Networks: Challenges and Future Directions}},
  author={Cook, Jonathan and Rehman, Sabih Ur and Khan, M Arif},
  journal={IEEE Access},
  volume={11},
  pages={39295--39317},
  year={2023},
  publisher={IEEE}
}

@article{pimentel2021understanding,
  title={{Understanding and improving the quality and reproducibility of Jupyter notebooks}},
  author={Pimentel, Jo{\~a}o Felipe and Murta, Leonardo and Braganholo, Vanessa and Freire, Juliana},
  journal={Empirical Software Engineering},
  volume={26},
  number={4},
  pages={2, {article no. 65}},
  year={2021},
  publisher={Springer}
}

@article{qiao2024bit,
  title={{Bit-Wise Mixture Differential Cryptanalysis and Its Application to SIMON}},
  author={Qiao, Kexin and Wu, Zehan and Cheng, Junjie and Ou, Changhai and Wang, An and Zhu, Liehuang},
  journal={IEEE Internet of Things Journal},
  year={2024},
  volume={11},
  number={13},
  pages={23398-23409},
  publisher={IEEE}
}

@misc{iso2018simon,
    author = {International Organization for Standardization},
    title = {{ISO/IEC 29167-21:2018(en)
Information technology — Automatic identification and data capture techniques — Part 21: Crypto suite SIMON security services for air interface communications}},
    howpublished = {\url{https://www.iso.org/obp/ui/es/#iso:std:iso-iec:29167:-21:ed-1:v1:en:sec:3.1.13}},
    year = {2018}
}

@book{wehrspohn2022take,
  title={{When Do I Take Which Distribution?: A Statistical Basis for Entrepreneurial Applications}},
  author={Wehrspohn, Uwe and Ernst, Dietmar},
  year={2022},
  pages={3-6},
  publisher={Springer Nature}
}

@inproceedings{biham1995matsui,
  title={{On Matsui's Linear Cryptanalysis}},
  author={Biham, Eli},
  booktitle={Advances in Cryptology—EUROCRYPT'94: Workshop on the Theory and Application of Cryptographic Techniques Perugia, Italy, May 9--12, 1994 Proceedings 13},
  pages={341--355},
  year={1995},
  organization={Springer}
}





% \end{thebibliography}

\end{document}